\documentclass[12pt]{article}
\usepackage[T2A]{fontenc}
\usepackage[english]{babel}

\usepackage{latexsym}
\usepackage{amssymb}
\usepackage{amsmath}

\def\proof{\medbreak\noindent{\bf Proof}}

\def\theorem #1. #2\par{\medbreak
  \noindent{\tt {\bf Theorem #1.}\enspace}{\sl#2\par}%
  \ifdim\lastskip<\medskipamount \removelastskip\penalty55\medskip\fi}

\def\lemma #1. #2\par{\medbreak
  \noindent{\tt {\bf Lemma #1.}\enspace}{\sl#2\par}%
  \ifdim\lastskip<\medskipamount \removelastskip\penalty55\medskip\fi}

\def\ch{{\rm ch}}
\def\sh{{\rm sh}}

\def\{{\lbrace}
\def\}{\rbrace}

\def\Wcl{W{\cal C}\!\ell}
\def\wcl{w{\cal C}\!\ell}

\def\cl{{\cal C}\!\ell}

\def\R{{\Bbb R}}
\def\C{{\Bbb C}}

\def\det{{\rm det}}

\def\exp{{\rm exp}}
\def\cos{{\rm cos}}
\def\sin{{\rm sin}}
\def\be{\begin{equation}}
\def\ee{\end{equation}}

\newcommand{\A}{{\cal A}}

\newcommand{\bbE}{{\mathbb E}}
\newcommand{\bbI}{{\mathbb I}}
\newcommand{\bbJ}{{\mathbb J}}
\newcommand{\bbK}{{\mathbb K}}
\newcommand{\st}{\stackrel}
\newcommand{\vsp}{{\vrule width0pt height15pt}}

\begin{document}
\title{Application of the method of quaternion typification for finding subalgebras and Lie subalgebras of Clifford algebras}

\author{D.~S.~Shirokov}
\maketitle
\begin{abstract}
In this paper we further develop the method of quaternion typification of Clifford algebra elements suggested by the author in the previous papers. On the basis of new classification of Clifford algebra elements it is possible to find out and prove a number of new properties of Clifford algebra. In particular, we find subalgebras and Lie subalgebras of Clifford algebra and subalgebras of the Lie algebra of the pseudo-unitary Lie group.
\end{abstract}

\tableofcontents

\bigskip
 In this paper we further develop the method of quaternion typification of Clifford algebra elements suggested by the author in the previous papers. On the basis of new classification of Clifford algebra elements it is possible to find out and prove a number of new properties of Clifford algebra. In particular, we find subalgebras and Lie subalgebras of Clifford algebra and subalgebras of the Lie algebra of the pseudo-unitary Lie group. We find all subalgebras in the form of linear combinations of elements of the given ranks or types.
We develop results of \cite{quattyp} and use results of \cite{Marchuk:Shirokov} and \cite{Shirokov}.

\bigskip


In the first section we shortly remind basics of the method of quaternion typification of Clifford algebra elements \cite{quattyp}.

\bigskip
\section{Main ideas of the method of quaternion typification of Clifford algebra elements}

Let $p, q$ be nonnegative integer numbers and $p+q=n$, $n\geq1$. Consider the real Clifford algebra $\cl^\R(p,q)$ or the complex Clifford algebra $\cl^\C(p,q)$. In the case when results are true for both cases, we write $\cl(p,q)$. The construction of Clifford algebra $\cl(p,q)$ is discussed in details in \cite{Lounesto} or \cite{Marchuk:Shirokov}. Let $e$ be the identity element and let $e^a$, $a=1,\ldots,n$ be generators of the Clifford algebra $\cl(p,q)$,
$$
e^a e^b+e^b e^a=2\eta^{ab}e,
$$
where $\eta=||\eta^{ab}||$ is the diagonal matrix with $p$ pieces of $+1$ and $q$ pieces of $-1$ on the diagonal. Elements
$$
e^{a_1\ldots a_k}=e^{a_1}\ldots e^{a_k},\quad a_1<\ldots<a_k,\,k=1,\ldots,n,
$$
together with the identity element $e$, form the basis of the Clifford
algebra. The number of basis elements is equal to $2^n$.

We denote by $\cl_k(p,q)$ the vector spaces that span over the basis elements
$e^{a_1\ldots a_k}$. Elements of $\cl_k(p,q)$ are said to be
elements of rank $k$. Sometimes we denote elements of rank $k$ by $\st{k}{W}, \st{k}{V}, \ldots$ We have the following classification of Clifford algebra elements based on the notion of rank:
\begin{eqnarray}
\cl(p,q)=\oplus_{k=0}^{n}\cl_k(p,q).\label{ranks}
\end{eqnarray}
So, any Clifford algebra element is an element of some rank or a sum of elements of different ranks:
\begin{eqnarray}
U=\st{k_1}{U}+\st{k_2}{U}+\ldots+\st{k_m}{U},\qquad 0\leq k_1<\ldots<k_m\leq n.
\end{eqnarray}

Also we have classification of Clifford algebra elements based on the notion of evenness:
\begin{eqnarray}
\cl(p,q)=\cl_{even}(p,q)\oplus\cl_{odd}(p,q),\label{evenness}
\end{eqnarray}
where
$$\cl_{even}(p,q)=\cl_0(p,q)\oplus\cl_2(p,q)\oplus\cl_4(p,q)\oplus\ldots,$$
$$\cl_{odd}(p,q)=\cl_1(p,q)\oplus\cl_3(p,q)\oplus\cl_5(p,q)\oplus\ldots$$

Any Clifford algebra element is an even element, an odd element or a sum of even and odd elements.

Denote by $[U,V]$ the commutator and  by $\{U,V\}$ the anticommutator of
Clifford algebra elements $U, V \in \cl(p,q)$
\begin{eqnarray}
[U,V]=UV-VU,\quad \{U,V\}=UV+VU \label{comanticom}
\end{eqnarray}
and note that
\begin{eqnarray}
UV=\frac{1}{2}[U,V]+\frac{1}{2}\{U,V\}.\label{proizv}
\end{eqnarray}

\bigskip

Consider the Clifford algebra as the vector space and represent it in the form of the direct sum of four subspaces:
\begin{equation}
\cl(p,q)=\cl_{\overline 0}(p,q)\oplus\cl_{\overline 1}(p,q)\oplus
\cl_{\overline 2}(p,q)\oplus\cl_{\overline 3}(p,q),\label{kv}
\end{equation}
where
\begin{eqnarray*}
\cl_{\overline
0}(p,q)&=&\cl_0(p,q)\oplus\cl_4(p,q)\oplus\cl_8(p,q)\oplus\ldots,\\
\cl_{\overline
1}(p,q)&=&\cl_1(p,q)\oplus\cl_5(p,q)\oplus\cl_9(p,q)\oplus\ldots,\\
\cl_{\overline
2}(p,q)&=&\cl_2(p,q)\oplus\cl_6(p,q)\oplus\cl_{10}(p,q)\oplus\ldots,\\
\cl_{\overline
3}(p,q)&=&\cl_3(p,q)\oplus\cl_7(p,q)\oplus\cl_{11}(p,q)\oplus\ldots
\end{eqnarray*}
and in the right hand parts there are direct sums of subspaces with dimensions differ on 4. We suppose that $\cl_k(p,q)=\emptyset$ for $k>p+q$.

If $\st{\overline k}{U}\in\cl_{\overline k}(p,q)$, then we have
$$
\st{\overline k}{U}=\st{k}{U}+\st{k+4}{U}+\st{k+8}{U}+\ldots, \qquad
k=0,1,2,3.
$$

We use the following notations:
$$
\cl_{\overline{kl}}(p,q)=\cl_{\overline
k}(p,q)\oplus\cl_{\overline l}(p,q),\quad 0\leq k<l\leq 3.
$$
$$
\cl_{\overline{klm}}(p,q)=\cl_{\overline
k}(p,q)\oplus\cl_{\overline l}(p,q)\oplus\cl_{\overline
m}(p,q),\quad 0\leq k<l<m\leq 3.
$$
If $\st{\overline{kl}}{U}\in\cl_{\overline{kl}}(p,q)$, then
$$
\st{\overline{kl}}{U}=\st{\overline{k}}{U}+\st{\overline{l}}{U}=
(\st{k}{U}+\st{l}{U})+(\st{k+4}{U}+\st{l+4}{U})+\ldots,\quad
0\leq k<l\leq 3.
$$

Consider elements of the Clifford algebra $\cl(p,q)$ from different subspaces
\begin{eqnarray}
\cl_{\overline 0}(p,q),\quad \cl_{\overline 1}(p,q),\quad \cl_{\overline 2}(p,q),\quad \cl_{\overline 3}(p,q),\quad \cl_{\overline {01}}(p,q),\quad \cl_{\overline {02}}(p,q),\nonumber \\
\cl_{\overline {03}}(p,q),\quad \cl_{\overline {12}}(p,q), \quad \cl_{\overline {13}}(p,q),\quad \cl_{\overline {23}}(p,q),\quad \cl_{\overline {012}}(p,q),\label{tip}\\
\cl_{\overline {013}}(p,q),\quad \cl_{\overline {023}}(p,q),\quad \cl_{\overline {123}}(p,q),\quad
\cl_{\overline {0123}}(p,q)=\cl(p,q).\nonumber
\end{eqnarray}
Then we say that these elements have different {\it quaternion types} (or {\it types}).

Elements of subspaces $\cl_{\overline 0}(p,q),\, \cl_{\overline 1}(p,q),\, \cl_{\overline 2}(p,q),\, \cl_{\overline 3}(p,q)$ are called {\it elements of the main quaternion types}. Elements of other types are represented in the form of sums of elements of the main quaternion types. Suppose that the zero element of the Clifford algebra $\cl(p,q)$ belongs to any quaternion type.

The classification of elements of the Clifford algebra $\cl(p,q)$ (for all integer nonnegative numbers $p+q=n$) on 15 quaternion types (see (\ref{tip})) and use statements of Theorem 1 (see \cite{quattyp}) for calculations of quaternion types of commutators and anticommutators of Clifford algebra elements is the essence of {\em the method of quaternion typification of Clifford algebra elements}.

\bigskip

Sometimes we denote subspace $\cl_{\overline k}(p,q)$ by $\overline{\textbf{k}}$ and any Clifford algebra element $\st{\overline k}{U}\in\cl_{\overline k}(p,q)$ by $\overline{k}$. When we write "quaternion type $\overline{k}$" we mean by $\overline{k}$ a symbol of quaternion type (not an Clifford algebra element). Then
$[\overline k,\overline l]\subseteq\overline{\textbf{m}}$ means that commutator of any two Clifford algebra elements of quaternion types $\overline k$ and $\overline l$ belongs to subspace $\overline{\textbf{m}}=\cl_{\overline m}(p,q)$. And $[\overline k,\overline l]=\overline{m}$ means that for any two Clifford algebra elements of quaternion types $\overline k$ ш $\overline l$ there exists a Clifford algebra element of quaternion type $\overline{m}$ and it equals to commutator.

Let's remind the definition of the algebra of quaternion type \cite{quattyp}:\\
Let $\A$ be an $n$-dimensional algebra over the field of complex or real numbers. And let algebra $\A$, considered as an $n$-dimensional vector space, be represented in the form of the direct sum of four vector subspaces
\begin{equation}
\A=\bbE\oplus\bbI\oplus\bbJ\oplus\bbK. \label{A}
\end{equation}
For the elements of these subspaces we use the following designations
$$
\st{\bbE}{A}\in\bbE,\quad\st{\bbI}{B}\in\bbI,\quad\st{\bbE\oplus\bbI}{C}
\in\bbE\oplus\bbI,\ldots
$$
An algebra $\A$ is called {\it the algebra of quaternion type with respect to an operation $\circ:\A\times\A\rightarrow\A$}, if for all elements of considered subspaces the following properties are fulfilled:
\begin{eqnarray}
&&\st{\bbE}{A}\circ\st{\bbE}{B},\ \st{\bbI}{A}\circ\st{\bbI}{B}, \
\st{\bbJ}{A}\circ\st{\bbJ}{B},\
\st{\bbK}{A}\circ\st{\bbK}{B}\in\bbE,\nonumber\\
&&\st{\bbE}{A}\circ\st{\bbI}{B},\ \st{\bbI}{A}\circ\st{\bbE}{B},\
\st{\bbK}{A}\circ\st{\bbJ}{B},\
\st{\bbJ}{A}\circ\st{\bbK}{B}\in\bbI,\label{q:cond}\\
&&\st{\bbE}{A}\circ\st{\bbJ}{B},\ \st{\bbJ}{A}\circ\st{\bbE}{B},\
\st{\bbI}{A}\circ\st{\bbK}{B},\ \st{\bbK}{A}\circ\st{\bbI}{B}\in\bbJ,\nonumber\\
&&\st{\bbE}{A}\circ\st{\bbK}{B},\ \st{\bbK}{A}\circ\st{\bbE}{B},\
\st{\bbI}{A}\circ\st{\bbJ}{B},\ \st{\bbJ}{A}\circ\st{\bbI}{B}\in\bbK.\nonumber
\end{eqnarray}
The operation $\circ$ unessentially should be associative or commutative.

From Theorem 1 \cite{quattyp} we have:\\
a) The Clifford algebra $\cl(p,q)$ is an algebra of quaternion type with respect to the operation $\quad U, V \rightarrow \{U,V\}$
and in this case
$$\quad \bbE=\cl_{\overline 0}(p,q),\quad\bbI=\cl_{\overline1}(p,q),\quad\bbJ=\cl_{\overline 2}(p,q),\quad\bbK=\cl_{\overline 3}(p,q)\quad.$$
b) The Clifford algebra $\cl(p,q)$ is an algebra of quaternion type with respect to the operation $\quad U, V \rightarrow [U,V]$
and in this case
$$\quad \bbE=\cl_{\overline 2}(p,q),\quad\bbI=\cl_{\overline 3}(p,q),\quad\bbJ=\cl_{\overline 0}(p,q),\quad\bbK=\cl_{\overline 1}(p,q)\quad.$$

These statements are equivalent to the following properties:
\begin{eqnarray}
&&[\st{\overline k}{U},\st{\overline k}{V}]=\st{\overline 2}{W},\qquad k=0, 1, 2, 3; \nonumber\\
&&[\st{\overline k}{U},\st{\overline 2}{V}]=\st{\overline k}{W}, \qquad k=0, 1, 2, 3; \nonumber \\
&&[\st{\overline 0}{U},\st{\overline 1}{V}]=\st{\overline 3}{W}, \quad  [\st{\overline 0}{U},\st{\overline 3}{V}]=\st{\overline 1}{W}, \quad [\st{\overline 1}{U},\st{\overline 3}{V}]=\st{\overline 0}{W} \nonumber.
\end{eqnarray}

\begin{eqnarray}
&&\{\st{\overline k}{U},\st{\overline k}{V}\}=\st{\overline 0}{W},\qquad k=0, 1, 2, 3; \nonumber\\
&&\{\st{\overline k}{U},\st{\overline 0}{V}\}=\st{\overline k}{W}, \qquad k=0, 1, 2, 3; \nonumber \\
&&\{\st{\overline 1}{U},\st{\overline 2}{V}\}=\st{\overline 3}{W},  \quad \{\st{\overline 1}{U},\st{\overline 3}{V}\}=\st{\overline 2}{W}, \quad \{\st{\overline 2}{U},\st{\overline 3}{V}\}=\st{\overline 1}{W}\nonumber.
\end{eqnarray}
Let's write down these and similar expressions in the other notation:
\begin{eqnarray}
&&[\overline k,\overline k]\subseteq\overline{\textbf{2}},\qquad k=0, 1, 2, 3; \nonumber\\
&&[\overline k,\overline 2]\subseteq\overline{\textbf{k}}, \qquad k=0, 1, 2, 3; \label{1} \\
&&[\overline 0,\overline 1]\subseteq\overline{\textbf{3}}, \quad  [\overline 0,\overline 3]\subseteq\overline{\textbf{1}}, \quad [\overline 1,\overline 3]\subseteq\overline{\textbf{0}} \nonumber,
\end{eqnarray}

\begin{eqnarray}
&&\{\overline k,\overline k\}\subseteq\overline{\textbf{0}},\qquad k=0, 1, 2, 3; \nonumber\\
&&\{\overline k,\overline 0\}\subseteq\overline{\textbf{k}}, \qquad k=0, 1, 2, 3; \label{2} \\
&&\{\overline 1,\overline 2\}\subseteq\overline{\textbf{3}},  \quad \{\overline 1,\overline 3\}\subseteq\overline{\textbf{2}}, \quad \{\overline 2,\overline 3\}\subseteq\overline{\textbf{1}}\nonumber.
\end{eqnarray}

\bigskip
The following tables display action of commutator, anticommutator and Clifford product of elements of the Clifford algebra of different quaternion types. By $\A$ denote the Clifford algebra $\cl(p,q)=\cl_{\overline 0\overline 1\overline 2\overline 3}(p,q)$.

\bigskip

\begin{tabular}{|p{0.5cm}|p{0.4cm} p{0.4cm} p{0.4cm} p{0.4cm} p{0.3cm} p{0.3cm} p{0.3cm} p{0.3cm} p{0.3cm} p{0.3cm} p{0.4cm} p{0.4cm} p{0.4cm} p{0.4cm} p{0.4cm}|}\hline\vsp
$[,]$ & $\overline 0$ & $\overline 1$ & $\overline 2$ & $\overline 3$ & $\overline 0\overline 1$ & $\overline 0\overline 2$ & $\overline 0\overline 3$ & $\overline 1\overline 2$ & $\overline 1\overline 3$ & $\overline 2\overline 3$ & $\overline 0\overline 1\overline 2$ & $\overline 0\overline 1\overline 3$ & $\overline 0\overline 2\overline 3$ & $\overline 1\overline 2\overline 3$ &$\A$\\ \hline\vsp
$\overline 0$ & $\overline 2$ & $\overline 3$ & $\overline 0$ & $\overline 1$ & $\overline 2\overline 3$ & $\overline 0\overline 2$ & $\overline 1\overline 2$ & $\overline 0\overline 3$ & $\overline 1\overline 3$ & $\overline 0\overline 1$ & $\overline 0\overline 2\overline 3$ & $\overline 1\overline 2\overline 3$ & $\overline 0\overline 1\overline 2$ & $\overline 0\overline 1\overline 3$ &$\A$\\
$\overline 1$ & $\overline3$ & $\overline 2$ & $\overline 1$ & $\overline 0$ & $\overline 2\overline 3$ & $\overline 1\overline 3$ & $\overline 0\overline 3$ & $\overline 1\overline 2$ & $\overline 0\overline 2$ & $\overline 0\overline 1$ & $\overline 1\overline 2\overline 3$ & $\overline 0\overline 2\overline 3$ & $\overline 0\overline 1\overline 3$ & $\overline 0\overline 1\overline 2$ &$\A$\\
$\overline 2$ & $\overline 0$ & $\overline 1$ & $\overline 2$ & $\overline 3$ & $\overline 0\overline 1$ & $\overline 0\overline 2$ & $\overline 0\overline 3$ & $\overline 1\overline 2$ & $\overline 1\overline 3$ & $\overline 2\overline 3$ & $\overline 0\overline 1\overline 2$ & $\overline 0\overline 1\overline 3$ & $\overline 0\overline 2\overline 3$ & $\overline 1\overline 2\overline 3$ &$\A$\\
$\overline 3$ & $\overline 1$ & $\overline 0$ & $\overline 3$ & $\overline 2$ & $\overline 0\overline 1$ & $\overline 1\overline 3$ & $\overline 1\overline 2$ & $\overline 0\overline 3$ & $\overline 0\overline 2$ & $\overline 2\overline 3$ & $\overline 0\overline 1\overline 3$ & $\overline 0\overline 1\overline 2$ & $\overline 1\overline 2\overline 3$ & $\overline 0\overline 2\overline 3$ &$\A$\\
$\overline 0\overline 1$ &$\overline 2\overline 3$ & $\overline 2\overline 3$ & $\overline 0\overline 1$ & $\overline 0\overline 1$ & $\overline 2\overline 3$ & $\A$ & $\A$ & $\A$ & $\A$ & $\overline 0\overline 1$ & $\A$ & $\A$ & $\A$ & $\A$ &$\A$\\
$\overline 0\overline 2$ & $\overline 0\overline2$ & $\overline 1\overline 3$ & $\overline 0\overline 2$ & $\overline 1\overline 3$ & $\A$ & $\overline 0\overline 2$ & $\A$ & $\A$ & $\overline 1\overline 3$ & $\A$ & $\A$ & $\A$ & $\A$ & $\A$ &$\A$\\
$\overline 0\overline 3$ & $\overline 1\overline2$ & $\overline 0\overline 3$ & $\overline 0\overline 3$ & $\overline 1\overline 2$ & $\A$ & $\A$ & $\overline 1\overline 2$ & $\overline 0\overline 3$ & $\A$ & $\A$ & $\A$ & $\A$ & $\A$ & $\A$ &$\A$\\
$\overline 1\overline 2$ & $\overline 0\overline3$ & $\overline 1\overline 2$ & $\overline 1\overline 2$ & $\overline 0\overline 3$ & $\A$ & $\A$ & $\overline 0\overline 3$ & $\overline 1\overline 2$ & $\A$ & $\A$ & $\A$ & $\A$ & $\A$ & $\A$ &$\A$\\
$\overline 1\overline 3$ & $\overline 1\overline3$ & $\overline 0\overline 2$ & $\overline 1\overline 3$ & $\overline 0\overline 2$ & $\A$ & $\overline 1\overline 3$ & $\A$ & $\A$ & $\overline 0\overline 2$ & $\A$ & $\A$ & $\A$ & $\A$ & $\A$ &$\A$\\
$\overline 2\overline 3$ & $\overline 0\overline 1$ & $\overline 0\overline 1$ & $\overline 2\overline 3$ & $\overline 2\overline 3$ & $\overline 0\overline 1$ & $\A$ & $\A$ & $\A$ & $\A$ & $\overline 2\overline 3$ & $\A$ & $\A$ & $\A$ & $\A$ &$\A$\\
$\overline 0\overline 1\overline 2$ & $\overline0\overline 2\overline 3$ & $\overline 1\overline 2\overline 3$ & $\overline 0\overline 1\overline 2$ & $\overline 0\overline 1\overline 3$ & $\A$ & $\A$ & $\A$ & $\A$ & $\A$ & $\A$ & $\A$ & $\A$ & $\A$ & $\A$ &$\A$\\
$\overline 0\overline 1\overline 3$ & $\overline1\overline 2\overline 3$ & $\overline 0\overline 2\overline 3$ & $\overline 0\overline 1\overline 3$ & $\overline 0\overline 1\overline 2$ & $\A$ & $\A$ & $\A$ & $\A$ & $\A$ & $\A$ & $\A$ & $\A$ & $\A$ & $\A$ &$\A$\\
$\overline 0\overline 2\overline 3$ & $\overline 0\overline 1\overline 2$ & $\overline 0\overline 1\overline 3$ & $\overline 0\overline 2\overline 3$ & $\overline 1\overline 2\overline 3$ & $\A$ & $\A$ & $\A$ & $\A$ & $\A$ & $\A$ & $\A$ & $\A$ & $\A$ & $\A$ &$\A$\\
$\overline 1\overline 2\overline 3$ & $\overline 0\overline 1\overline 3$ & $\overline 0\overline 1\overline 2$ & $\overline 1\overline 2\overline 3$ & $\overline 0\overline 2\overline 3$ & $\A$ & $\A$ & $\A$ & $\A$ & $\A$ & $\A$ & $\A$ & $\A$ & $\A$ & $\A$ &$\A$\\
$\A$ & $\A$ & $\A$ & $\A$ & $\A$ & $\A$ & $\A$ & $\A$ & $\A$ & $\A$ & $\A$ & $\A$ & $\A$ & $\A$ & $\A$ & $ \A$ \\ \hline
\end{tabular}
\bigskip

\begin{tabular}{|p{0.5cm}|p{0.4cm} p{0.4cm} p{0.4cm} p{0.4cm} p{0.3cm} p{0.3cm} p{0.3cm} p{0.3cm} p{0.3cm} p{0.3cm} p{0.4cm} p{0.4cm} p{0.4cm} p{0.4cm} p{0.4cm}|}\hline\vsp
$\{,\}$ & $\overline 0$ & $\overline 1$ & $\overline 2$ & $\overline 3$ & $\overline 0\overline 1$ & $\overline 0\overline 2$ & $\overline 0\overline 3$ & $\overline 1\overline 2$ & $\overline 1\overline 3$ & $\overline 2\overline 3$ & $\overline 0\overline 1\overline 2$ & $\overline 0\overline 1\overline 3$ & $\overline 0\overline 2\overline 3$ & $\overline 1\overline 2\overline 3$ &$\A$\\ \hline\vsp
$\overline 0$ & $\overline 0$ & $\overline 1$ & $\overline 2$ & $\overline 3$ & $\overline 0\overline 1$ & $\overline 0\overline 2$ & $\overline 0\overline 3$ & $\overline 1\overline 2$ & $\overline 1\overline 3$ & $\overline 2\overline 3$ & $\overline 0\overline 1\overline 2$ & $\overline 0\overline 1\overline 3$ & $\overline 0\overline 2\overline 3$ & $\overline 1\overline 2\overline 3$ &$\A$\\
$\overline 1$ & $\overline 1$ & $\overline 0$ & $\overline 3$ & $\overline 2$ & $\overline 0\overline 1$ & $\overline 1\overline 3$ & $\overline 1\overline 2$ & $\overline 0\overline 3$ & $\overline 0\overline 2$ & $\overline 2\overline 3$ & $\overline 0\overline 1\overline 3$ & $\overline 0\overline 1\overline 2$ & $\overline 1\overline 2\overline 3$ & $\overline 0\overline 2\overline 3$ &$\A$\\
$\overline 2$ & $\overline 2$ & $\overline 3$ & $\overline 0$ & $\overline 1$ & $\overline 2\overline 3$ & $\overline 0\overline 2$ & $\overline 1\overline 2$ & $\overline 0\overline 3$ & $\overline 1\overline 3$ & $\overline 0\overline 1$ & $\overline 0\overline 2\overline 3$ & $\overline 1\overline 2\overline 3$ & $\overline 0\overline 1\overline 2$ & $\overline 0\overline 1\overline 3$ &$\A$\\
$\overline 3$ & $\overline 3$ & $\overline 2$ & $\overline 1$ & $\overline 0$ & $\overline 2\overline 3$ & $\overline 1\overline 3$ & $\overline 0\overline 3$ & $\overline 1\overline 2$ & $\overline 0\overline 2$ & $\overline 0\overline 1$ & $\overline 1\overline 2\overline 3$ & $\overline 0\overline 2\overline 3$ & $\overline 0\overline 1\overline 3$ & $\overline 0\overline 1\overline 2$ &$\A$\\
$\overline 0\overline 1$ & $\overline 0\overline 1$ & $\overline 0\overline 1$ & $\overline 2\overline 3$ & $\overline 2\overline 3$ & $\overline 0\overline 1$ & $\A$ & $\A$ & $\A$ & $\A$ & $\overline 2\overline 3$ & $\A$ & $\A$ & $\A$ & $\A$ &$\A$\\
$\overline 0\overline 2$ & $\overline 0\overline 2$ & $\overline 1\overline 3$ & $\overline 0\overline 2$ & $\overline 1\overline 3$ & $\A$ & $\overline 0\overline 2$ & $\A$ & $\A$ & $\overline 1\overline 3$ & $\A$ & $\A$ & $\A$ & $\A$ & $\A$ &$\A$\\
$\overline 0\overline 3$ & $\overline 0\overline 3$ & $\overline 1\overline 2$ & $\overline 1\overline 2$ & $\overline 0\overline 3$ & $\A$ & $\A$ & $\overline 0\overline 3$ & $\overline 1\overline 2$ & $\A$ & $\A$ & $\A$ & $\A$ & $\A$ & $\A$ &$\A$\\
$\overline 1\overline 2$ & $\overline 1\overline 2$ & $\overline 0\overline 3$ & $\overline 0\overline 3$ & $\overline 1\overline 2$ & $\A$ & $\A$ & $\overline 1\overline 2$ & $\overline 0\overline 3$ & $\A$ & $\A$ & $\A$ & $\A$ & $\A$ & $\A$ &$\A$\\
$\overline 1\overline 3$ & $\overline 1\overline 3$ & $\overline 0\overline 2$ & $\overline 1\overline 3$ & $\overline 0\overline 2$ & $\A$ & $\overline 1\overline 3$ & $\A$ & $\A$ & $\overline 0\overline 2$ & $\A$ & $\A$ & $\A$ & $\A$ & $\A$ &$\A$\\
$\overline 2\overline 3$ & $\overline 2\overline 3$ & $\overline 2\overline 3$ & $\overline 0\overline 1$ & $\overline 0\overline 1$ & $\overline 2\overline 3$ & $\A$ & $\A$ & $\A$ & $\A$ & $\overline 0\overline 1$ & $\A$ & $\A$ & $\A$ & $\A$ &$\A$\\
$\overline 0\overline 1\overline 2$ & $\overline 0\overline 1\overline 2$ & $\overline 0\overline 1\overline 3$ & $\overline 0\overline 2\overline 3$ & $\overline 1\overline 2\overline 3$ & $\A$ & $\A$ & $\A$ & $\A$ & $\A$ & $\A$ & $\A$ & $\A$ & $\A$ & $\A$ &$\A$\\
$\overline 0\overline 1\overline 3$ & $\overline 0\overline 1\overline 3$ & $\overline 0\overline 1\overline 2$ & $\overline 1\overline 2\overline 3$ & $\overline 0\overline 2\overline 3$ & $\A$ & $\A$ & $\A$ & $\A$ & $\A$ & $\A$ & $\A$ & $\A$ & $\A$ & $\A$ &$\A$\\
$\overline 0\overline 2\overline 3$ & $\overline 0\overline 2\overline 3$ & $\overline 1\overline 2\overline 3$ & $\overline 0\overline 1\overline 2$ & $\overline 0\overline 1\overline 3$ & $\A$ & $\A$ & $\A$ & $\A$ & $\A$ & $\A$ & $\A$ & $\A$ & $\A$ & $\A$ &$\A$\\
$\overline 1\overline 2\overline 3$ & $\overline 1\overline 2\overline 3$ & $\overline 0\overline 2\overline 3$ & $\overline 0\overline 1\overline 3$ & $\overline 0\overline 1\overline 2$ & $\A$ & $\A$ & $\A$ & $\A$ & $\A$ & $\A$ & $\A$ & $\A$ & $\A$ & $\A$ &$\A$\\
$\A$ & $\A$ & $\A$ & $\A$ & $\A$ & $\A$ & $\A$ & $\A$ & $\A$ & $\A$ & $\A$ & $\A$ & $\A$ & $\A$ & $\A$ & $ \A$ \\ \hline
\end{tabular}

\bigskip

\begin{tabular}{|p{0.5cm}|p{0.4cm} p{0.4cm} p{0.4cm} p{0.4cm} p{0.3cm} p{0.3cm} p{0.3cm} p{0.3cm} p{0.3cm} p{0.3cm} p{0.4cm} p{0.4cm} p{0.4cm} p{0.4cm} p{0.4cm}|}\hline\vsp
& $\overline 0$ & $\overline 1$ & $\overline 2$ & $\overline 3$ & $\overline 0\overline 1$ & $\overline 0\overline 2$ & $\overline 0\overline 3$ & $\overline 1\overline 2$ & $\overline 1\overline 3$ & $\overline 2\overline 3$ & $\overline 0\overline 1\overline 2$ & $\overline 0\overline 1\overline 3$ & $\overline 0\overline 2\overline 3$ & $\overline 1\overline 2\overline 3$ &$\A$\\ \hline\vsp
$\overline 0$ & $\overline {02}$ & $\overline {13}$ & $\overline {02}$ & $\overline {13}$ & $\A$ & $\overline {02}$ & $\A$ & $\A$ & $\overline {13}$ & $\A$ & $\A$ & $\A$ & $\A$ & $\A$ &$\A$\\
$\overline 1$ & $\overline {13}$ & $\overline {02}$ & $\overline {13}$ & $\overline {02}$ & $\A$ & $\overline {13}$ & $\A$ & $\A$ & $\overline {02}$ & $\A$ & $\A$ & $\A$ & $\A$ & $\A$ &$\A$\\
$\overline 2$ & $\overline {02}$ & $\overline {13}$ & $\overline {02}$ & $\overline {13}$ & $\A$ & $\overline {02}$ & $\A$ & $\A$ & $\overline {13}$ & $\A$ & $\A$ & $\A$ & $\A$ & $\A$ &$\A$\\
$\overline 3$ & $\overline {13}$ & $\overline {02}$ & $\overline {13}$ & $\overline {02}$ & $\A$ & $\overline {13}$ & $\A$ & $\A$ & $\overline {02}$ & $\A$ & $\A$ & $\A$ & $\A$ & $\A$ &$\A$\\
$\overline {01}$ & $\A$ & $A$ & $\A$ & $\A$ & $\A$ & $\A$ & $\A$ & $\A$ & $\A$ & $\A$ & $\A$ & $\A$ & $\A$ & $\A$ &$\A$\\
$\overline {02}$ & $\overline {02}$ & $\overline {13}$ & $\overline {02}$ & $\overline {13}$ & $\A$ & $\overline {02}$ & $\A$ & $\A$ & $\overline {13}$ & $\A$ & $\A$ & $\A$ & $\A$ & $\A$ &$\A$\\
$\overline {03}$ & $\A$ & $\A$ & $\A$ & $\A$ & $\A$ & $\A$ & $\A$ & $\A$ & $\A$ & $\A$ & $\A$ & $\A$ & $\A$ & $\A$ &$\A$\\
$\overline {12}$ & $\A$ & $\A$ & $\A$ & $\A$ & $\A$ & $\A$ & $\A$ & $\A$ & $\A$ & $\A$ & $\A$ & $\A$ & $\A$ & $\A$ &$\A$\\
$\overline {13}$ & $\overline {13}$ & $\overline {02}$ & $\overline {13}$ & $\overline {02}$ & $\A$ & $\overline {13}$ & $\A$ & $\A$ & $\overline {02}$ & $\A$ & $\A$ & $\A$ & $\A$ & $\A$ &$\A$\\
$\overline {23}$ & $\A$ & $\A$ & $\A$ & $\A$ & $\A$ & $\A$ & $\A$ & $\A$ & $\A$ & $\A$ & $\A$ & $\A$ & $\A$ & $\A$ &$\A$\\
$\overline {012}$ & $\A$ & $\A$ & $\A$ & $\A$ & $\A$ & $\A$ & $\A$ & $\A$ & $\A$ & $\A$ & $\A$ & $\A$ & $\A$ & $\A$ &$\A$\\
$\overline {013}$ & $\A$ & $\A$ & $\A$ & $\A$ & $\A$ & $\A$ & $\A$ & $\A$ & $\A$ & $\A$ & $\A$ & $\A$ & $\A$ & $\A$ &$\A$\\
$\overline {023}$ & $\A$ & $\A$ & $\A$ & $\A$ & $\A$ & $\A$ & $\A$ & $\A$ & $\A$ & $\A$ & $\A$ & $\A$ & $\A$ & $\A$ &$\A$\\
$\overline {123}$ & $\A$ & $\A$ & $\A$ & $\A$ & $\A$ & $\A$ & $\A$ & $\A$ & $\A$ & $\A$ & $\A$ & $\A$ & $\A$ & $\A$ &$\A$\\
$\A$ & $\A$ & $\A$ & $\A$ & $\A$ & $\A$ & $\A$ & $\A$ & $\A$ & $\A$ & $\A$ & $\A$ & $\A$ & $\A$ & $\A$ & $ \A$ \\ \hline
\end{tabular}

\bigskip

\section{Subalgebras in the form of linear combinations of elements of the given types}
The method of quaternion typification of Clifford algebra elements allow us to prove a number of new properties of Clifford algebras.

In this section we denote $\cl_{\overline{k}}^\R(p,q)$ by $\,\overline{\textbf{k}}\,$ and $\cl_{\overline{k}}^\C(p,q)$ by $\,\overline{\textbf{k}}\oplus i\overline{\textbf{k}}\,$.

\begin{theorem}1.
a) The subspace
\begin{equation}
\overline{\textbf{02}}=\cl^\R_{even}(p,q)
\end{equation}
forms subalgebra of the real Clifford algebra $\cl^\R(p,q)$.\\
b) Subspaces
\begin{align}
\overline{\textbf{02}}&=\cl^\R_{even}(p,q),& \overline{\textbf{02}}\oplus i\overline{\textbf{02}}&=\cl^\C_{even}(p,q) ,\\
\overline{\textbf{02}}\oplus i\overline{\textbf{13}}&=\cl^\R_{even}(p,q)\oplus i\cl^\R_{odd}(p,q),& \overline{\textbf{0123}}&=\cl^\R(p,q)\nonumber
\end{align}
form subalgebras of the complex Clifford algebra $\cl^\C(p,q)$.\\

\end{theorem}

\proof. \, With the aid of written out above table the proof of this theorem is straightforward. $\blacksquare$

\begin{theorem}2.
a) Subspaces
\begin{equation}
\overline{\textbf{2}},\qquad\overline{\textbf{02}},\qquad \overline{\textbf{12}},\qquad \overline{\textbf{23}}
\end{equation}
of the real Clifford algebra $\cl^\R(p,q)$ are closed with respect
to the commutator $\quad U, V \rightarrow [U,V]$ and, hence, form Lie algebras w.r.t. the commutator.\\
b) Subspaces
\begin{eqnarray}
&&\overline{\textbf{2}},\qquad\overline{\textbf{02}},\qquad\overline{\textbf{12}},\qquad\overline{\textbf{23}},\qquad\overline{\textbf{0123}},\nonumber\\
&&\overline{\textbf{02}}\oplus i\overline{\textbf{02}},\qquad\overline{\textbf{12}}\oplus i\overline{\textbf{12}},\qquad\overline{\textbf{23}}\oplus i\overline{\textbf{23}},\label{subalg}\\
&&\overline{\textbf{2}}\oplus i\overline{\textbf{0}},\qquad\overline{\textbf{2}}\oplus i\overline{\textbf{1}},\qquad\overline{\textbf{2}}\oplus i\overline{\textbf{2}},\qquad\overline{\textbf{2}}\oplus i\overline{\textbf{3}},\nonumber\\
&&\overline{\textbf{02}}\oplus i\overline{\textbf{13}},\qquad\overline{\textbf{12}}\oplus i\overline{\textbf{03}},\qquad\overline{\textbf{23}}\oplus i\overline{\textbf{01}}\nonumber
\end{eqnarray}
of the complex Clifford algebra $\cl^\C(p,q)$ are closed with respect
to the commutator $\quad U, V \rightarrow [U,V]$ and, hence, form Lie algebras w.r.t the commutator.\\

\end{theorem}

\begin{theorem}3.
a) Subspaces
\begin{equation}
\overline{\textbf{0}},\qquad\overline{\textbf{01}},\qquad \overline{\textbf{02}},\qquad \overline{\textbf{03}}
\end{equation}
of the real Clifford algebra $\cl^\R(p,q)$ are closed with respect to the operation $\quad U, V \rightarrow \{U,V\}$ and form subalgebras of the Clifford algebra considered with respect to the operation $\quad U, V \rightarrow \{U,V\}$.\\
b) Subspaces
\begin{eqnarray}
&&\overline{\textbf{0}},\qquad\overline{\textbf{01}},\qquad\overline{\textbf{02}},\qquad\overline{\textbf{03}},\qquad\overline{\textbf{0123}},\nonumber\\
&&\overline{\textbf{01}}\oplus i\overline{\textbf{01}},\qquad\overline{\textbf{02}}\oplus i\overline{\textbf{02}},\qquad\overline{\textbf{03}}\oplus i\overline{\textbf{03}},\label{subalg2}\\
&&\overline{\textbf{0}}\oplus i\overline{\textbf{0}},\qquad\overline{\textbf{0}}\oplus i\overline{\textbf{1}},\qquad\overline{\textbf{0}}\oplus i\overline{\textbf{2}},\qquad\overline{\textbf{0}}\oplus i\overline{\textbf{3}},\nonumber\\
&&\overline{\textbf{01}}\oplus i\overline{\textbf{23}},\qquad\overline{\textbf{02}}\oplus i\overline{\textbf{13}},\qquad\overline{\textbf{03}}\oplus i\overline{\textbf{12}}\nonumber
\end{eqnarray}
of the complex Clifford algebra $\cl^\C(p,q)$ are closed with
respect to the anticommutator $\quad U, V \rightarrow \{U,V\}$ and
form subalgebras of the Clifford algebra considered with respect
to the operation $\quad U, V \rightarrow \{U,V\}$.

\end{theorem}

\proof. \, With the aid of (\ref{1}),(\ref{2}) (or see above tables) the proof of this theorem is straightforward.$\blacksquare$

\bigskip
Now we consider the notions of the pseudo-unitary group $\Wcl^\C(p,q)$ of the complex Clifford algebra and the Lie algebra $\wcl^\C(p,q)$ of the Lie group $\Wcl^\C(p,q)$ (see in \cite{Shirokov}).

Consider the following set of Clifford algebra elements:
\begin{equation}
\Wcl^\C(p,q)=\{U\in\cl^\C(p,q): U^* U=e\}, \label{psungr}
\end{equation}
where * is the operation of Clifford conjugation
\cite{Marchuk:Shirokov} with properties
$$
e^*=e,\quad (e^a)^*=e^a,\quad (\lambda\ e^{a_1}e^{a_2}\ldots e^{a_k})^*=\overline{\lambda}\ e^{a_k}\ldots e^{a_1},\quad
$$
$\lambda$ is a complex number and $\overline{\lambda}$ is the
conjugated complex number. This set forms a (Lie) group with respect
to the Clifford product and this group is called {\it the
pseudo-unitary group of the Clifford algebra $\cl(p,q)$ }.

The set of elements with the commutator $[U,V]=UV-VU$
\begin{equation}
\wcl^\C(p,q)=\{u\in\cl^\C(p,q): u^*=-u\}.\label{liealg}
\end{equation}
is {\it the Lie algebra of the Lie group $\Wcl^\C(p,q)$}.

From this definition and from the definition of Clifford conjugation  it follows that an arbitrary element of this Lie algebra has the form
$$
u=i\st{0}{u}+i\st{1}{u}+\st{2}{u}+\st{3}{u}+i\st{4}{u}+i\st{5}{u}+\ldots+
a_n\st{n}{u}=\sum_{k=0}^{n}a_k \st{k}{u},
$$
where $\st{k}{u} \in\cl_k^\R (p,q)$ and
$$
a_k=\left\lbrace
\begin{array}{ll}
1, & \mbox{\rm $k=2, 3, 6, 7, \ldots$;}\\
i, & \mbox{\rm $k=0, 1, 4, 5, \ldots$}
\end{array}
\right.
$$

So
\begin{equation}
\wcl^\C(p,q)=i\cl_{\overline{0}}^\R(p,q) \oplus i\cl_{\overline{1}}^\R(p,q) \oplus \cl_{\overline{2}}^\R(p,q) \oplus \cl_{\overline 3}^\R(p,q).\label{wcl}
\end{equation}

\begin{theorem}4. The Lie algebra $\wcl^\C(p,q)$ of the Lie group $\Wcl^\C(p,q)$ is an algebra of quaternion type with respect to the operation $\quad U, V \rightarrow [U,V]$ and $$\quad\bbE=\cl_{\overline{2}}^\R(p,q),\quad\bbI=\cl_{\overline{3}}^\R(p,q),\quad\bbJ=i\cl_{\overline{0}}^\R(p,q),\quad\bbK=i\cl_{\overline 1}^\R(p,q)\quad.$$

\end{theorem}

\proof. \, The statement of the theorem is equivalent to the following properties:
\begin{align}
[i\overline k,i\overline k]&\subseteq\overline{\textbf{2}},& k&=0, 1, \nonumber\\
[\overline k,\overline k]&\subseteq\overline{\textbf{2}},& k&=2, 3, \nonumber\\
[i\overline k,\overline 2]&\subseteq i\overline{\textbf{k}},& k&=0, 1, \label{th4}\\
[\overline k,\overline 2]&\subseteq\overline{\textbf{k}},& k&=3, \nonumber \\
[i\overline 0,i\overline 1]&\subseteq\overline{\textbf{3}},\quad [i\overline 0,\overline 3]\subseteq i\overline{\textbf{1}},\quad[i\overline 1,\overline 3]\subseteq i\overline{\textbf{0}} \nonumber.
\end{align}
But these formulas follow from (\ref{1}). These completes the proof of the theorem. $\blacksquare$

\begin{theorem}5.
Subspaces
\begin{equation}
\overline{\textbf{2}},\qquad\overline{\textbf{2}}\oplus i\overline{\textbf{0}},\qquad\overline{\textbf{2}}\oplus i\overline{\textbf{1}},\qquad\overline{\textbf{23}}
\end{equation}
of the complex Clifford algebra $\cl^\C(p,q)$ are closed with respect to the operation $\quad U, V \rightarrow [U,V]$ and form subalgebras of the Lie algebra $\wcl^\C(p,q)$ of the pseudo-unitary group of the Clifford algebra.\\

\end{theorem}

\proof. \, With the aid of (\ref{subalg}) and (\ref{wcl}) the proof of this theorem is straightforward.$\blacksquare$

\begin{theorem}6.
The following subspaces form subgroups of pseudo-unitary group $\Wcl^\C(p,q)$. The Lie algebras from Theorem 5 correspond to these Lie groups.\\
\bigskip
$$\begin{tabular}{|c|c|}\hline
Lie algebra & Lie group\\\hline
$\qquad\overline{\textbf{2}}\qquad$&$\{U\in\overline{\textbf{02}}=\cl^\R_{even}(p,q): U^* U=e\}$\\\hline
$\qquad\overline{\textbf{2}}\oplus i\overline{\textbf{0}}\qquad$&$\{U\in\overline{\textbf{02}}\oplus i\overline{\textbf{02}}=\cl^\C_{even}(p,q): U^* U=e\}$\\\hline
$\qquad\overline{\textbf{2}}\oplus i\overline{\textbf{1}}\qquad$&$\{U\in\overline{\textbf{02}}\oplus i\overline{\textbf{13}}=\cl^\R_{even}(p,q)\oplus i\cl^\R_{odd}(p,q): U^* U=e\}$\\\hline
$\qquad\overline{\textbf{23}}\qquad$&$\{U\in\overline{\textbf{0123}}=\cl^\R(p,q): U^* U=e\}$\\\hline
\end{tabular}$$
\bigskip

\end{theorem}

\proof. \, Let's prove, for example, the first of four statements. Let $U$ be an element of Lie group $\{U\in\overline{\textbf{02}}: U^* U=e\}$. Then
\begin{equation}
U=e+\varepsilon u,
\end{equation}
where $\varepsilon^2=0$ and $u$ - an element of the real Lie algebra of this Lie group (there is only one such Lie algebra). Then
\begin{equation}
e=U^*U=(e+\varepsilon u^*)(e+\varepsilon u)=e+\varepsilon(u+u^*).\nonumber
\end{equation}
So, for element of Lie algebra we have $u^*=-u$, i.e. $u\in\overline{\textbf{23}}\oplus i\overline{\textbf{01}}$. But also $u\in\overline{\textbf{02}}$. Thus, $u\in\overline{\textbf{2}}$.$\blacksquare$

\section{Subalgebras in the form of linear combinations of elements of the given ranks}
Note that classification of Clifford algebra elements based on the notion of quaternion type is rougher than the classification based on the notion of rank. So, let's discuss our problem in detail. In this section we search subalgebras and Lie subalgebras in the form of linear combinations of elements of the given ranks.

\begin{theorem}7.
Let $\st{k}{U}, \st{l}{V}, \st{r}{W}$ be Clifford algebra $\cl(p,q)$ elements of ranks $k, l$ and $r$.
Then, for all integer $n\geq k\geq l\geq 0$ we have
\begin{equation}
\st{k}{U}\st{l}{V}=\left\lbrace
\begin{array}{ll}
\st{k-l}{W}+\st{k-l+2}{W}+\ldots+\st{k+l}{W}, & \mbox{\rm $k+l\leq n$;}\\
\st{k-l}{W}+\st{k-l+2}{W}+\ldots+\st{2n-k-l}{W}, & \mbox{\rm $k+l\geq n$.}
\end{array}
\right.
\label{hest}
\end{equation}

\end{theorem}

\proof. \, This statement follows from Theorems 1 and 2 from \cite{Shirokov}.$\blacksquare$
Note that this theorem makes more exact theorem from \cite{Hestenes}:
$$\st{k}{U}\st{l}{V}=\st{k-l}{W}+\st{k-l+2}{W}+\ldots+\st{k+l}{W}, \quad \mbox{уфх}\st{m}{W}=0\quad \mbox{for} \quad m>n,\quad \mbox{and}\quad m<0$$

Let denote $\cl_k^\R(p,q)$ by $\widehat{\textbf{k}}$.
\begin{theorem}8.
a) Subspaces
\begin{eqnarray}
&&\widehat{\textbf{0}},\quad \widehat{\textbf{0}}\oplus\widehat{\textbf{n}},\quad\widehat{\textbf{0}}\oplus\widehat{\textbf{2}}\oplus\ldots\oplus\widehat{\textbf{k}}=\overline{\textbf{02}},\quad k=n, n-1
\end{eqnarray}
form subalgebras of the real Clifford algebra $\cl^\R(p,q)$.\\
b) Subspaces
\begin{eqnarray}
&&\widehat{\textbf{0}},\quad \widehat{\textbf{0}}\oplus\widehat{\textbf{n}},\nonumber\\
&&\widehat{\textbf{0}}\oplus\widehat{\textbf{2}}\oplus\ldots\oplus\widehat{\textbf{k}}=\overline{\textbf{02}},\quad k=n, n-1\nonumber\\
&&\widehat{\textbf{0}}\oplus i\widehat{\textbf{0}},\quad \widehat{\textbf{0}}\oplus\widehat{\textbf{n}}\oplus i\widehat{\textbf{0}}\oplus i\widehat{\textbf{n}},\\
&&\widehat{\textbf{0}}\oplus i\widehat{\textbf{0}}\oplus \widehat{\textbf{2}}\oplus i\widehat{\textbf{2}}\ldots\oplus\widehat{\textbf{k}}\oplus i\widehat{\textbf{k}}=\overline{\textbf{02}}\oplus i\overline{\textbf{02}},\quad k=n, n-1\nonumber\\
&&\widehat{\textbf{0}}\oplus i\widehat{\textbf{1}}\oplus \widehat{\textbf{2}}\oplus i\widehat{\textbf{3}}\oplus\ldots i^k(-1)^{k(k-1)/2}\widehat{\textbf{n}}=\overline{\textbf{02}}\oplus i \overline{\textbf{13}}\nonumber\\
&&\widehat{\textbf{0}}\oplus \widehat{\textbf{1}}\oplus \widehat{\textbf{2}}\oplus\ldots\oplus\widehat{\textbf{n}}=\overline{\textbf{0123}}\nonumber
\end{eqnarray}
form subalgebras of the complex Clifford algebra $\cl^\C(p,q)$.\\

\end{theorem}

\proof. \, With the aid of Theorem 7 the proof of this theorem is straightforward.$\blacksquare$

Let
$$
a_k=\left\lbrace
\begin{array}{ll}
1, & \mbox{\rm $k=2, 3, 6, 7, \ldots$;}\\
i, & \mbox{\rm $k=0, 1, 4, 5, \ldots$}
\end{array}
\right.
$$
Note that classification of subalgebras is conventional in the following theorems. This gradation help us to orientate in great number of subalgebras.

\begin{theorem}9.
The following subspaces of the real Clifford algebra $\cl^\R(p,q)$ are closed with respect to the commutator $\quad U, V \rightarrow [U,V]$ and, hence, form Lie algebras:
\begin{description}
\item 1) for $n\geq 1$: $$\widehat{\textbf{0}};$$
\item 2) for $n\geq 1$: $$\widehat{\textbf{n}};$$
\item 3) for $n\geq 2$: $$\widehat{\textbf{1}}\oplus\widehat{\textbf{2}};$$
\item 4) for $n\geq 3$ (if $n=2$ it is the same as item 2): $$\widehat{\textbf{2}};$$
\item 5) for $n\geq 4$ (if $n=2, 3$ it is the same as item 3): $$\widehat{\textbf{1}}\oplus\widehat{\textbf{2}}\oplus\ldots\oplus \widehat{\textbf{n}}$$ for even n, $$\widehat{\textbf{1}}\oplus \widehat{\textbf{2}}\oplus \ldots\oplus \widehat{\textbf{n-1}}$$ for odd n;
\item 6) for $n\geq 4$: $$\widehat{\textbf{2}}\oplus\widehat{\textbf{n-1}};$$
\item 7) for $n\geq 5$: $$\widehat{\textbf{2}}\oplus\widehat{\textbf{n-2}};$$
\item 8) for $n\geq 6$ (if $n=5$ it is the same as item 5): $$\widehat{\textbf{1}}\oplus \widehat{\textbf{2}}\oplus \widehat{\textbf{n-2}}\oplus \widehat{\textbf{n-1}}$$ for odd n , $$\widehat{\textbf{1}}\oplus \widehat{\textbf{2}}\oplus \widehat{\textbf{n-1}}\oplus \widehat{\textbf{n}}$$ for even n;
\item 9) for $n\geq 6$ (if $n=2, 3$ it is the same as item 4, if $n=4$ it is the same as item 6, if $n=5$ it is the same as item 7):$$\widehat{\textbf{2}}\oplus \widehat{\textbf{3}}\oplus \widehat{\textbf{6}}\oplus \widehat{\textbf{7}}\oplus \widehat{\textbf{10}}\oplus \widehat{\textbf{11}}\oplus \ldots\oplus
\widehat{\textbf{k}}=\overline{\textbf{23}}$$ for $n=k+1, k+2$ for odd $k$ and $n=k, k+1$ for even $k$;
\item 10) for $n\geq 7$ (if $n=3, 4$ it is the same as item 4, if $n=5$ it is the same as item 6, if $n=6$ it is the same as item 7): $$\widehat{\textbf{2}}\oplus \widehat{\textbf{4}}\oplus \widehat{\textbf{6}}\oplus \widehat{\textbf{8}}\oplus \widehat{\textbf{10}}\oplus \widehat{\textbf{12}}\oplus \ldots\oplus
\widehat{\textbf{k}}=\overline {\textbf{02}}$$ for $n=k+1, k+2$;
\item 11) for $n\geq 8$ (if $n=2, 3, 4, 5$ it is the same as item 3, if $n=6, 7$ it is the same as item 8): $$\widehat{\textbf{1}}\oplus \widehat{\textbf{2}}\oplus \widehat{\textbf{5}}\oplus \widehat{\textbf{6}}\oplus \widehat{\textbf{9}}\oplus \widehat{\textbf{10}}\oplus \ldots\oplus
\widehat{\textbf{k}}=\overline {\textbf{12}}$$ for $n=k, k+1, k+2, k+3$ for even $k$;
\item 12) for $n\geq 9$ (if $n=3, 4, 5, 6$ it is the same as item 4, if $n=7$ it is the same as item 6, if $n=8$ it is the same as item 7): $$\widehat{\textbf{2}}\oplus \widehat{\textbf{6}}\oplus \widehat{\textbf{10}}\oplus \widehat{\textbf{14}}\oplus \widehat{\textbf{18}}\oplus \widehat{\textbf{22}}\oplus \ldots\oplus \widehat{\textbf{k}}=\overline{\textbf{2}}$$ for $n=k+1, k+2, k+3, k+4$.
\end{description}
(In all items equality to subspaces of quaternion types are understood to within an element of rank $0$ and rank $n$).\\
Besides, the direct sums of all listed subalgebras with $\widehat{\textbf{0}}$ are also Lie subalgebras for any $n$. The direct sums of all listed subalgebras with $\widehat{\textbf{n}}$ are Lie subalgebras for odd $n$. Also we can add $\widehat{\textbf{n}}$ to subalgebras that consist of elements of even ranks for even $n$.
(These cases aren't in the 1)-12) items of the theorem because we get reducible subalgebras.)

\end{theorem}

\begin{theorem}10.
The following subspaces of the complex Clifford algebra $\cl^\C(p,q)$ are closed with respect to the commutator $\quad U, V \rightarrow [U,V]$ and, hence, form Lie algebras:
\begin{description}
\item 1.1-1.3) for $n\geq 1$: $$\widehat{\textbf{0}},$$ $$i\widehat{\textbf{0}},$$ $$\widehat{\textbf{0}}\oplus i\widehat{\textbf{0}};$$
\item 2.1-2.3) for $n\geq 1$: $$\widehat{\textbf{n}},$$ $$i\widehat{\textbf{n}},$$ $$\widehat{\textbf{n}}\oplus i\widehat{\textbf{n}};$$
\item 3.1-3.3) for $n\geq 2$: $$\widehat{\textbf{1}}\oplus\widehat{\textbf{2}},$$ $$i\widehat{\textbf{1}}\oplus\widehat{\textbf{2}},$$ $$\widehat{\textbf{1}}\oplus\widehat{\textbf{2}}\oplus i \widehat{\textbf{1}}\oplus i\widehat{\textbf{2}};$$
\item 4.1-4.2) for $n\geq 3$ (if $n=2$ it is the same as item 2): $$\widehat{\textbf{2}},$$ $$\widehat{\textbf{2}}\oplus i\widehat{\textbf{2}};$$
\item 5.1-5.5) for $n\geq 4$ (if $n=2, 3$ it is the same as item 3): $$\widehat{\textbf{1}}\oplus\widehat{\textbf{2}}\oplus\ldots\oplus \widehat{\textbf{n}}=\overline{\textbf{0123}},$$ $$i\widehat{\textbf{1}}\oplus\widehat{\textbf{2}}\oplus\ldots\oplus a_n\widehat{\textbf{n}}=\overline{\textbf{23}}\oplus i\overline{\textbf{01}},$$ $$i\widehat{\textbf{1}}\oplus\widehat{\textbf{2}}\oplus\ldots\oplus i^n (-1)^{n(n-1)/2}\widehat{\textbf{n}}=\overline{\textbf{02}}\oplus i\overline{\textbf{13}},$$ $$\widehat{\textbf{1}}\oplus\widehat{\textbf{2}}\oplus\ldots\oplus a_{n+1}\widehat{\textbf{n}}=\overline{\textbf{12}}\oplus i\overline{\textbf{03}},$$ $$\widehat{\textbf{1}}\oplus\widehat{\textbf{2}}\oplus\ldots\oplus \widehat{\textbf{n}}\oplus i\widehat{\textbf{1}}\oplus i\widehat{\textbf{2}}\oplus\ldots\oplus i \widehat{\textbf{n}} =\overline{\textbf{0123}}\oplus i\overline{\textbf{0123}}$$for even $n$; $$\widehat{\textbf{1}}\oplus \widehat{\textbf{2}}\oplus \ldots\oplus \widehat{\textbf{n-1}}=\overline{{0123}},$$ $$i\widehat{\textbf{1}}\oplus \widehat{\textbf{2}}\oplus \ldots\oplus a_{n-1}\widehat{\textbf{n-1}}=\overline{\textbf{23}}\oplus i\overline{\textbf{01}},$$ $$i\widehat{\textbf{1}}\oplus \widehat{\textbf{2}}\oplus \ldots\oplus i^{n-1}(-1)^{(n-1)(n-2)/2}\widehat{\textbf{n-1}}=\overline{\textbf{02}}\oplus i\overline{\textbf{13}},$$ $$\widehat{\textbf{1}}\oplus \widehat{\textbf{2}}\oplus \ldots\oplus a_{n}\widehat{\textbf{n-1}}=\overline{\textbf{12}}\oplus i\overline{\textbf{03}},$$  $$\widehat{\textbf{1}}\oplus \widehat{\textbf{2}}\oplus \ldots\oplus \widehat{\textbf{n-1}}\oplus i\widehat{\textbf{1}}\oplus i\widehat{\textbf{2}}\oplus \ldots\oplus i\widehat{\textbf{n-1}} =\overline{\textbf{0123}}\oplus i\overline{\textbf{0123}}$$ for odd $n$;
\item 6.1-6.3) for $n\geq 4$: $$\widehat{\textbf{2}}\oplus\widehat{\textbf{n-1}},$$ $$ \widehat{\textbf{2}}\oplus\widehat{\textbf{n-1}}\oplus i\widehat{\textbf{2}}\oplus i\widehat{\textbf{n-1}},$$ $$\widehat{\textbf{2}}\oplus i \widehat{\textbf{n-1}};$$
\item 7.1-7.3) for $n\geq 5$: $$\widehat{\textbf{2}}\oplus\widehat{\textbf{n-2}},$$ $$ \widehat{\textbf{2}}\oplus\widehat{\textbf{n-2}}\oplus i\widehat{\textbf{2}}\oplus i\widehat{\textbf{n-2}},$$ $$\widehat{\textbf{2}}\oplus i \widehat{\textbf{n-2}};$$
\item 8.1-8.5) for $n\geq 6$ (if $n=5$ it is the same as item 5): $$\widehat{\textbf{1}}\oplus \widehat{\textbf{2}}\oplus \widehat{\textbf{n-2}}\oplus \widehat{\textbf{n-1}},$$ $$ \widehat{\textbf{1}}\oplus \widehat{\textbf{2}}\oplus \widehat{\textbf{n-2}}\oplus \widehat{\textbf{n-1}}\oplus i\widehat{\textbf{1}}\oplus i\widehat{\textbf{2}}\oplus i\widehat{\textbf{n-2}}\oplus i\widehat{\textbf{n-1}},$$ $$\widehat{\textbf{1}}\oplus \widehat{\textbf{2}}\oplus i\widehat{\textbf{n-2}}\oplus i\widehat{\textbf{n-1}},$$ $$i\widehat{\textbf{1}}\oplus \widehat{\textbf{2}}\oplus \widehat{\textbf{n-2}}\oplus i\widehat{\textbf{n-1}},$$ $$i\widehat{\textbf{1}}\oplus \widehat{\textbf{2}}\oplus i\widehat{\textbf{n-2}}\oplus \widehat{\textbf{n-1}}$$ for odd $n$ ; $$\widehat{\textbf{1}}\oplus \widehat{\textbf{2}}\oplus \widehat{\textbf{n-1}}\oplus \widehat{\textbf{n}},$$ $$ \widehat{\textbf{1}}\oplus \widehat{\textbf{2}}\oplus \widehat{\textbf{n-1}}\oplus \widehat{\textbf{n}}\oplus i\widehat{\textbf{1}}\oplus i\widehat{\textbf{2}}\oplus i\widehat{\textbf{n-1}}\oplus i\widehat{\textbf{n}},$$ $$\widehat{\textbf{1}}\oplus \widehat{\textbf{2}}\oplus i\widehat{\textbf{n-1}}\oplus i\widehat{\textbf{n}},$$ $$i\widehat{\textbf{1}}\oplus \widehat{\textbf{2}}\oplus \widehat{\textbf{n-1}}\oplus i\widehat{\textbf{n}},$$ $$i\widehat{\textbf{1}}\oplus \widehat{\textbf{2}}\oplus i\widehat{\textbf{n-1}}\oplus \widehat{\textbf{n}}$$ for even $n$;
\item 9.1-9.3) for $n\geq 6$ (if $n=2, 3$ it is the same as item 4, if $n=4$ it is the same as item 6, if $n=5$ it is the same as item 7):$$\widehat{\textbf{2}}\oplus \widehat{\textbf{3}}\oplus \widehat{\textbf{6}}\oplus \widehat{\textbf{7}}\oplus \widehat{\textbf{10}}\oplus \widehat{\textbf{11}}\oplus \ldots\oplus
\widehat{\textbf{k}}=\overline{\textbf{23}},$$ $$\widehat{\textbf{2}}\oplus i\widehat{\textbf{3}}\oplus \widehat{\textbf{6}}\oplus i\widehat{\textbf{7}}\oplus \widehat{\textbf{10}}\oplus i\widehat{\textbf{11}}\oplus \ldots\oplus
a_{k+2}\widehat{\textbf{k}}=\overline{\textbf{2}}\oplus i \overline{\textbf{3}},$$ $$\widehat{\textbf{2}}\oplus \widehat{\textbf{3}}\oplus \widehat{\textbf{6}}\oplus \widehat{\textbf{7}}\oplus \ldots\oplus
\widehat{\textbf{k}}\oplus i\widehat{\textbf{2}}\oplus i\widehat{\textbf{3}}\oplus i\widehat{\textbf{6}}\oplus i\widehat{\textbf{7}}\oplus \ldots\oplus
i\widehat{\textbf{k}}=\overline{\textbf{23}}\oplus i\overline{\textbf{23}}$$ for $n=k+1, k+2$ for odd $k$ and $n=k, k+1$ for even $k$;
\item 10.1-10.3) for $n\geq 7$ (if $n=3, 4$ it is the same as item 4, if $n=5$ it is the same as item 6, if $n=6$ it is the same as item 7): $$\widehat{\textbf{2}}\oplus \widehat{\textbf{4}}\oplus \widehat{\textbf{6}}\oplus \widehat{\textbf{8}}\oplus \widehat{\textbf{10}}\oplus \widehat{\textbf{12}}\oplus \ldots\oplus
\widehat{\textbf{k}}=\overline {\textbf{02}},$$ $$\widehat{\textbf{2}}\oplus i\widehat{\textbf{4}}\oplus \widehat{\textbf{6}}\oplus i\widehat{\textbf{8}}\oplus \widehat{\textbf{10}}\oplus i\widehat{\textbf{12}}\oplus \ldots\oplus
a_{k}\widehat{\textbf{k}}=\overline{\textbf{2}}\oplus i\overline{\textbf{0}}$$ $$ \widehat{\textbf{2}}\oplus \widehat{\textbf{4}}\oplus \widehat{\textbf{6}}\oplus \widehat{\textbf{8}}\oplus \ldots\oplus
\widehat{\textbf{k}}\oplus i\widehat{\textbf{2}}\oplus i\widehat{\textbf{4}}\oplus i\widehat{\textbf{6}}\oplus i\widehat{\textbf{8}}\oplus \ldots\oplus
i\widehat{\textbf{k}}=\overline {\textbf{02}}\oplus i \overline{\textbf{02}}$$ for $n=k+1, k+2$;
\item 11.1-11.3) for $n\geq 8$ (if $n=2, 3, 4, 5$ it is the same as item 3, if $n=6, 7$ it is the same as item 8): $$\widehat{\textbf{1}}\oplus \widehat{\textbf{2}}\oplus \widehat{\textbf{5}}\oplus \widehat{\textbf{6}}\oplus \widehat{\textbf{9}}\oplus \widehat{\textbf{10}}\oplus \ldots\oplus \widehat{\textbf{k}}=\overline {\textbf{12}},$$ $$i\widehat{\textbf{1}}\oplus \widehat{\textbf{2}}\oplus i\widehat{\textbf{5}}\oplus \widehat{\textbf{6}}\oplus i\widehat{\textbf{9}}\oplus \widehat{\textbf{10}}\oplus \ldots\oplus
a_{k}\widehat{\textbf{k}}=\overline{\textbf{2}}\oplus i\overline{\textbf{1}}$$ $$ \widehat{\textbf{1}}\oplus \widehat{\textbf{2}}\oplus \widehat{\textbf{5}}\oplus \widehat{\textbf{6}}\oplus \ldots\oplus \widehat{\textbf{k}}\oplus i\widehat{\textbf{1}}\oplus i\widehat{\textbf{2}}\oplus i\widehat{\textbf{5}}\oplus i\widehat{\textbf{6}}\oplus \ldots\oplus i\widehat{\textbf{k}}=\overline {\textbf{12}}\oplus i \overline{\textbf{12}}$$ for $n=k, k+1, k+2, k+3$ for even $k$;
\item 12.1-12.2) for $n\geq 9$ (if $n=3, 4, 5, 6$ it is the same as item 4, if $n=7$ it is the same as item 6, if $n=8$ it is the same as item 7): $$\widehat{\textbf{2}}\oplus \widehat{\textbf{6}}\oplus \widehat{\textbf{10}}\oplus \widehat{\textbf{14}}\oplus \widehat{\textbf{18}}\oplus \widehat{\textbf{22}}\oplus \ldots\oplus \widehat{\textbf{k}}=\overline{\textbf{2}},$$ $$ \widehat{\textbf{2}}\oplus \widehat{\textbf{6}}\oplus \widehat{\textbf{10}}\oplus \widehat{\textbf{14}}\oplus \ldots\oplus \widehat{\textbf{k}}\oplus i\widehat{\textbf{2}}\oplus i\widehat{\textbf{6}}\oplus i\widehat{\textbf{10}}\oplus i\widehat{\textbf{14}}\oplus \ldots\oplus i\widehat{\textbf{k}}=\overline{\textbf{2}}\oplus i \overline{\textbf{2}}$$ for $n=k+1, k+2, k+3, k+4$.
\end{description}
(In all items equality to subspaces of quaternion types are understood to within an element of rank $0$ or rank $n$).\\
Besides, the direct sums of all listed subalgebras with $\widehat{\textbf{0}}, i\widehat{\textbf{0}}$ are also Lie subalgebras for any $n$. The direct sums of all listed subalgebras with $\widehat{\textbf{n}}, i\widehat{\textbf{n}}$ are Lie subalgebras for odd $n$. Also we can add $\widehat{\textbf{n}}, i\widehat{\textbf{n}}$ to subalgebras that consist of elements of even ranks for even $n$.
(These cases aren't in the 1)-12) items of the theorem because we get reducible subalgebras.)

\end{theorem}

\begin{theorem}11.
The following subspaces of the real Clifford algebra $\cl^\R(p,q)$ are closed with respect to the anticommutator $\quad U, V \rightarrow \{U,V\}$ and, hence, form subalgebras of the Clifford algebra considered with respect to the operation $\quad U, V \rightarrow \{U,V\}$:
\begin{description}
\item 1) for $n\geq 1$: $$\widehat{\textbf{0}};$$
\item 2) for $n\geq 2$: $$\widehat{\textbf{0}}\oplus\widehat{\textbf{1}};$$
\item 3) for $n\geq 2$: $$\widehat{\textbf{0}}\oplus\widehat{\textbf{n}};$$
\item 4) for $n\geq 3$: $$\widehat{\textbf{0}}\oplus\widehat{\textbf{n-1}};$$
\item 5) for $n\geq 4$: $$\widehat{\textbf{0}}\oplus\widehat{\textbf{1}}\oplus\widehat{\textbf{n}}$$ for even $n$;
\item 6) for $n\geq 4$: $$\widehat{\textbf{0}}\oplus\widehat{\textbf{n-1}}\oplus\widehat{\textbf{n}}$$ for even $n$;
\item 7) for $n\geq 4$ (if $n=1$ it is the same as item 1, if $n=2$ it is the same as item 3, if $n=3$ it is the same as item 4):$$\widehat{\textbf{0}}\oplus \widehat{\textbf{2}}\oplus \widehat{\textbf{4}}\oplus \widehat{\textbf{6}}\oplus \widehat{\textbf{8}}\oplus \widehat {\textbf{10}}\oplus \ldots\oplus\widehat{\textbf{k}}=\overline{\textbf{02}}$$ for $n=k+1, k+2$;
\item 8) for $n\geq 5$: $$\widehat{\textbf{0}}\oplus \widehat{\textbf{1}}\oplus \widehat{\textbf{n-1}}\oplus \widehat{\textbf{n}}$$ for odd $n$;
\item 9) for $n\geq 5$ (if $n=1, 2$ it is the same as item 1, if $n=3$ it is the same as item 3, if $n=4$ it is the same as item 6): $$\widehat{\textbf{0}}\oplus \widehat{\textbf{3}}\oplus \widehat{\textbf{4}}\oplus \widehat{\textbf{7}}\oplus \widehat{\textbf{8}}\oplus \widehat{\textbf{11}}\oplus \ldots\oplus
\widehat{\textbf{k}}=\overline{\textbf{03}}$$ for $n=k, k+1, k+2$ for even $k$ and $n=k$ for odd $k$;
\item 10) for $n\geq 6$ (if $n=2, 3$ it is the same as item 2, if $n=4$ it is the same as item 5, if $n=5$ it is the same as item 8): $$\widehat{\textbf{0}}\oplus \widehat{\textbf{1}}\oplus \widehat{\textbf{4}}\oplus \widehat{\textbf{5}}\oplus \widehat{\textbf{8}}\oplus \widehat{\textbf{9}}\oplus \ldots\oplus
\widehat{\textbf{k}}=\overline {\textbf{01}}$$ for $n=k, k+1, k+2$ for odd $k$ and $n=k$ for even $k$;
\item 11) for $n\geq 6$ (if $n=1, 2, 3$ it is the same as item 1, if $n=4$ it is the same as item 3, if $n=5$ it is the same as item 4): $$\widehat{\textbf{0}}\oplus \widehat{\textbf{4}}\oplus \widehat{\textbf{8}}\oplus \widehat{\textbf{12}}\oplus \widehat{\textbf{16}}\oplus \widehat{\textbf{20}}\oplus \ldots\oplus \widehat{\textbf{k}}=\overline{\textbf{0}}$$ for $n=k, k+1, k+2, k+3$.
\end{description}

\end{theorem}

\begin{theorem}12.
The following subspaces of the complex Clifford algebra $\cl^\C(p,q)$ are closed with respect to the anticommutator $\quad U, V \rightarrow \{U,V\}$ and, hence, form subalgebras of the Clifford algebra considered with respect to the operation $\quad U, V \rightarrow \{U,V\}$:
\begin{description}
\item 1.1-1.2) for $n\geq 1$: $$\widehat{\textbf{0}},$$ $$\widehat{\textbf{0}}\oplus i\widehat{\textbf{0}};$$
\item 2.1-2.3) for $n\geq 2$: $$\widehat{\textbf{0}}\oplus\widehat{\textbf{1}},$$ $$\widehat{\textbf{0}}\oplus i\widehat{\textbf{1}},$$ $$\widehat{\textbf{0}}\oplus\widehat{\textbf{1}}\oplus i\widehat{\textbf{0}}\oplus i\widehat{\textbf{1}};$$
\item 3.1-3.3) for $n\geq 2$: $$\widehat{\textbf{0}}\oplus\widehat{\textbf{n}},$$ $$\widehat{\textbf{0}}\oplus i\widehat{\textbf{n}},$$ $$\widehat{\textbf{0}}\oplus\widehat{\textbf{n}}\oplus i\widehat{\textbf{0}}\oplus i\widehat{\textbf{n}};$$
\item 4.1-4.3) for $n\geq 3$: $$\widehat{\textbf{0}}\oplus\widehat{\textbf{n-1}},$$ $$\widehat{\textbf{0}}\oplus i\widehat{\textbf{n-1}},$$ $$\widehat{\textbf{0}}\oplus\widehat{\textbf{n-1}}\oplus i\widehat{\textbf{0}}\oplus i\widehat{\textbf{n-1}};$$
\item 5.1-5.5) for $n\geq 4$: $$\widehat{\textbf{0}}\oplus\widehat{\textbf{1}}\oplus\widehat{\textbf{n}},$$ $$\widehat{\textbf{0}}\oplus i\widehat{\textbf{1}}\oplus i\widehat{\textbf{n}},$$ $$\widehat{\textbf{0}}\oplus i\widehat{\textbf{1}}\oplus\widehat{\textbf{n}},$$ $$\widehat{\textbf{0}}\oplus\widehat{\textbf{1}}\oplus i\widehat{\textbf{n}},$$ $$\widehat{\textbf{0}}\oplus\widehat{\textbf{1}}\oplus\widehat{\textbf{n}}\oplus i\widehat{\textbf{0}}\oplus i\widehat{\textbf{1}}\oplus i\widehat{\textbf{n}}$$ for even $n$;
\item 6.1-6.5) for $n\geq 4$: $$\widehat{\textbf{0}}\oplus\widehat{\textbf{n-1}}\oplus\widehat{\textbf{n}},$$ $$\widehat{\textbf{0}}\oplus i\widehat{\textbf{n-1}}\oplus i\widehat{\textbf{n}},$$ $$\widehat{\textbf{0}}\oplus i\widehat{\textbf{n-1}}\oplus\widehat{\textbf{n}},$$ $$\widehat{\textbf{0}}\oplus\widehat{\textbf{n-1}}\oplus i\widehat{\textbf{n}},$$ $$\widehat{\textbf{0}}\oplus\widehat{\textbf{n-1}}\oplus\widehat{\textbf{n}}\oplus i\widehat{\textbf{0}}\oplus i\widehat{\textbf{n-1}}\oplus i\widehat{\textbf{n}}$$ for even $n$;
\item 7.1-7.3) for $n\geq 4$ (if $n=1$ it is the same as item 1, if $n=2$ it is the same as item 3, if $n=3$ it is the same as item 4):$$\widehat{\textbf{0}}\oplus \widehat{\textbf{2}}\oplus \widehat{\textbf{4}}\oplus \widehat{\textbf{6}}\oplus \widehat{\textbf{8}}\oplus \widehat{\textbf{10}}\oplus \ldots\oplus\widehat{\textbf{k}}=\overline{\textbf{02}},$$ $$\widehat{\textbf{0}}\oplus i\widehat{\textbf{2}}\oplus \widehat{\textbf{4}}\oplus i\widehat{\textbf{6}}\oplus \widehat{\textbf{8}}\oplus i\widehat{\textbf{10}}\oplus \ldots\oplus i a_k\widehat{\textbf{k}}=\overline{\textbf{0}}\oplus i\overline{\textbf{2}},$$ $$\widehat{\textbf{0}}\oplus \widehat{\textbf{2}}\oplus \widehat{\textbf{4}}\oplus \widehat{\textbf{6}}\oplus \ldots\oplus\widehat{\textbf{k}}\oplus i\widehat{\textbf{0}}\oplus i\widehat{\textbf{2}}\oplus i\widehat{\textbf{4}}\oplus i\widehat{\textbf{6}}\oplus \ldots\oplus i\widehat{\textbf{k}}=\overline{\textbf{02}}\oplus i\overline{\textbf{02}}$$ фы  $n=k+1, k+2$;
\item 8.1-8.5) for $n\geq 5$: $$\widehat{\textbf{0}}\oplus \widehat{\textbf{1}}\oplus \widehat{\textbf{n-1}}\oplus \widehat{\textbf{n}},$$ $$\widehat{\textbf{0}}\oplus \widehat{\textbf{1}}\oplus i\widehat{\textbf{n-1}}\oplus i\widehat{\textbf{n}},$$ $$\widehat{\textbf{0}}\oplus i\widehat{\textbf{1}}\oplus i\widehat{\textbf{n-1}}\oplus \widehat{\textbf{n}},$$ $$\widehat{\textbf{0}}\oplus i\widehat{\textbf{1}}\oplus \widehat{\textbf{n-1}}\oplus i\widehat{\textbf{n}},$$ $$\widehat{\textbf{0}}\oplus \widehat{\textbf{1}}\oplus \widehat{\textbf{n-1}}\oplus \widehat{\textbf{n}}\oplus i\widehat{\textbf{0}}\oplus i\widehat{\textbf{1}}\oplus i \widehat{\textbf{n-1}}\oplus i\widehat{\textbf{n}}$$ for odd n;
\item 9.1-9.3) for $n\geq 5$ (if $n=1, 2$ it is the same as item 1, if $n=3$ it is the same as item 3, if $n=4$ it is the same as item 6): $$\widehat{\textbf{0}}\oplus \widehat{\textbf{3}}\oplus \widehat{\textbf{4}}\oplus \widehat{\textbf{7}}\oplus \widehat{\textbf{8}}\oplus \widehat{\textbf{11}}\oplus \ldots\oplus
\widehat{\textbf{k}}=\overline{\textbf{03}},$$ $$\widehat{\textbf{0}}\oplus i\widehat{\textbf{3}}\oplus \widehat{\textbf{4}}\oplus i\widehat{\textbf{7}}\oplus \widehat{\textbf{8}}\oplus i\widehat{\textbf{11}}\oplus \ldots\oplus
ia_k\widehat{\textbf{k}}=\overline{\textbf{0}}\oplus i\overline{\textbf{3}},$$ $$\widehat{\textbf{0}}\oplus \widehat{\textbf{3}}\oplus \widehat{\textbf{4}}\oplus \widehat{\textbf{7}}\oplus \ldots\oplus
\widehat{\textbf{k}}\oplus i\widehat{\textbf{0}}\oplus i\widehat{\textbf{3}}\oplus i\widehat{\textbf{4}}\oplus i\widehat{\textbf{7}}\oplus \ldots\oplus
i\widehat{\textbf{k}}=\overline{\textbf{03}}\oplus i\overline{\textbf{03}}$$ for $n=k, k+1, k+2$ for even $k$ and $n=k$ for odd $k$;
\item 10.1-10.3) for $n\geq 6$ (if $n=2, 3$ it is the same as item 2, if $n=4$ it is the same as item 5, if $n=5$ it is the same as item 8): $$\widehat{\textbf{0}}\oplus \widehat{\textbf{1}}\oplus \widehat{\textbf{4}}\oplus \widehat{\textbf{5}}\oplus \widehat{\textbf{8}}\oplus \widehat{\textbf{9}}\oplus \ldots\oplus
\widehat{\textbf{k}}=\overline{\textbf{01}},$$ $$\widehat{\textbf{0}}\oplus i\widehat{\textbf{1}}\oplus \widehat{\textbf{4}}\oplus i\widehat{\textbf{5}}\oplus \widehat{\textbf{8}}\oplus i\widehat{\textbf{9}}\oplus \ldots\oplus
a_{k-1}\widehat{\textbf{k}}=\overline{\textbf{0}}\oplus i\overline{\textbf{1}},$$ $$\widehat{\textbf{0}}\oplus \widehat{\textbf{1}}\oplus \widehat{\textbf{4}}\oplus \widehat{\textbf{5}}\oplus \ldots\oplus
\widehat{\textbf{k}}\oplus i\widehat{\textbf{0}}\oplus i\widehat{\textbf{1}}\oplus i\widehat{\textbf{4}}\oplus i\widehat{\textbf{5}}\oplus \ldots\oplus
i\widehat{\textbf{k}}=\overline{\textbf{01}}\oplus i\overline{\textbf{01}}$$ for $n=k, k+1, k+2$ for odd $k$ and $n=k$ for even $k$;
\item 11.1-11.2) for $n\geq 6$ (if $n=1, 2, 3$ it is the same as item 1, if $n=4$ it is the same as item 3, if $n=5$ it is the same as item 4): $$\widehat{\textbf{0}}\oplus \widehat{\textbf{4}}\oplus \widehat{\textbf{8}}\oplus \widehat{\textbf{12}}\oplus \widehat{\textbf{16}}\oplus \widehat{\textbf{20}}\oplus \ldots\oplus \widehat{\textbf{k}}=\overline{\textbf{0}},$$ $$\widehat{\textbf{0}}\oplus \widehat{\textbf{4}}\oplus \widehat{\textbf{8}}\oplus \widehat{\textbf{12}}\oplus \ldots\oplus \widehat{\textbf{k}}\oplus i\widehat{\textbf{0}}\oplus i\widehat{\textbf{4}}\oplus i\widehat{\textbf{8}}\oplus i\widehat{\textbf{12}}\oplus \ldots\oplus i\widehat{\textbf{k}}=\overline{\textbf{0}}\oplus i\overline{\textbf{0}}$$ for $n=k, k+1, k+2, k+3$;
\item 12.1-12.4) for $n\geq 2$: $$\widehat{\textbf{0}}\oplus \widehat{\textbf{1}}\oplus\widehat{\textbf{2}}\oplus\ldots\oplus \widehat{\textbf{n}}=\overline{\textbf{0123}},$$ $$\widehat{\textbf{0}}\oplus \widehat{\textbf{1}}\oplus i\widehat{\textbf{2}}\oplus\ldots\oplus ia_n\widehat{\textbf{n}}=\overline{\textbf{01}}\oplus i\overline{\textbf{23}},$$ $$\widehat{\textbf{0}}\oplus i\widehat{\textbf{1}}\oplus\widehat{\textbf{2}}\oplus\ldots\oplus i^n (-1)^{n(n-1)/2}\widehat{\textbf{n}}=\overline{\textbf{02}}\oplus i\overline{\textbf{13}},$$ $$\widehat{\textbf{0}}\oplus i\widehat{\textbf{1}}\oplus i\widehat{\textbf{2}}\oplus\ldots\oplus a_{n-1}\widehat{\textbf{n}}=\overline{\textbf{03}}\oplus i\overline{\textbf{12}}$$.
\end{description}

\end{theorem}

\proof. \, Proof of Theorems 9, 10, 11, 12 follows from the statements of Theorems 1 and 2 from \cite{Shirokov}.$\blacksquare$

Now let's speak about Lie subalgebras of the Lie algebra $\wcl(p,q)$ of pseudo-unitary group of Clifford algebra $\cl^\C(p,q)$. This result can be found in \cite{Shirokov}.
\begin{theorem}13.
The following subspaces of Clifford algebra $\cl^\C(p,q)$ form subalgebras of Lie algebra $\wcl(p,q)$:
\begin{description}
\item 1) for $n\geq 1$: $$i\widehat{\textbf{0}};$$
\item 2) for $n\geq 1$: $$a_{n}\widehat{\textbf{n}};$$
\item 3) for $n\geq 2$: $$i\widehat{\textbf{1}}\oplus\widehat{\textbf{2}};$$
\item 4) for $n\geq 3$ (if $n=2$ it is the same as item 2): $$\widehat{\textbf{2}};$$
\item 5) for $n\geq 4$ (if $n=2, 3$ it is the same as item 3): $$i\widehat{\textbf{1}}\oplus\widehat{\textbf{2}}\oplus\ldots\oplus a_{n}\widehat{\textbf{n}}$$ for even $n$, $$i\widehat{\textbf{1}}\oplus \widehat{\textbf{2}}\oplus \ldots\oplus a_{n-1}\widehat{\textbf{n-1}}$$ for odd $n$;
\item 6) for $n\geq 4$: $$\widehat{\textbf{2}}\oplus a_{n-1}\widehat{\textbf{n-1}};$$
\item 7) for $n\geq 5$: $$\widehat{\textbf{2}}\oplus a_{n-2}\widehat{\textbf{n-2}};$$
\item 8) for $n\geq 6$ if $n=5$ it is the same as item 5): $$i\widehat{\textbf{1}}\oplus \widehat{\textbf{2}}\oplus a_{n-2}\widehat{\textbf{n-2}}\oplus a_{n-1}\widehat{\textbf{n-1}}$$ for odd $n$ , $$i\widehat{\textbf{1}}\oplus \widehat{\textbf{2}}\oplus a_{n-1}\widehat{\textbf{n-1}}\oplus a_{n}\widehat{\textbf{n}}$$ for even $n$;
\item 9) for $n\geq 6$ (if $n=2, 3$ it is the same as item 4, if $n=4$ it is the same as item 6, if $n=5$ it is the same as item 7):$$\widehat{\textbf{2}}\oplus \widehat{\textbf{3}}\oplus \widehat{\textbf{6}}\oplus \widehat{\textbf{7}}\oplus \widehat{\textbf{10}}\oplus \widehat{\textbf{11}}\oplus \ldots\oplus
\widehat{\textbf{k}}=\overline{\textbf{23}}$$ for $n=k+1, k+2$ for odd $k$ and $n=k, k+1$ for even $k$;
\item 10) for $n\geq 7$ (if $n=3, 4$ it is the same as item 4, if $n=5$ it is the same as item 6, if $n=6$ it is the same as item 7): $$\widehat{\textbf{2}}\oplus i\widehat{\textbf{4}}\oplus \widehat{\textbf{6}}\oplus i\widehat{\textbf{8}}\oplus \widehat{\textbf{10}}\oplus i\widehat{\textbf{12}}\oplus \ldots\oplus
a_{k}\widehat{\textbf{k}}=\overline{\textbf{2}}\oplus i\overline{\textbf{0}}$$ фы  $n=k+1, k+2$;
\item 11) for $n\geq 8$ (if $n=2, 3, 4, 5$ it is the same as item 3, if $n=6, 7$ it is the same as item 8): $$i\widehat{\textbf{1}}\oplus \widehat{\textbf{2}}\oplus i\widehat{\textbf{5}}\oplus \widehat{\textbf{6}}\oplus i\widehat{\textbf{9}}\oplus \widehat{\textbf{10}}\oplus \ldots\oplus
a_{k}\widehat{\textbf{k}}=\overline{\textbf{2}}\oplus i\overline{\textbf{1}}$$ for $n=k, k+1, k+2, k+3$ for even $k$;
\item 12) for $n\geq 9$ (if $n=3, 4, 5, 6$ it is the same as item 4, if $n=7$ it is the same as item 6, if $n=8$ it is the same as item 7): $$\widehat{\textbf{2}}\oplus \widehat{\textbf{6}}\oplus \widehat{\textbf{10}}\oplus \widehat{\textbf{14}}\oplus \widehat{\textbf{18}}\oplus \widehat{\textbf{22}}\oplus \ldots\oplus \widehat{\textbf{k}}=\overline{\textbf{2}}$$ for $n=k+1, k+2, k+3, k+4$.
\end{description}
Besides, the direct sums of all listed subalgebras with $i\widehat{\textbf{0}}$ are also Lie subalgebras for any $n$. The direct sums of all listed subalgebras with  $a_n\widehat{\textbf{n}}$ are Lie subalgebras for odd $n$. Also we can add $a_n\widehat{\textbf{n}}$ to subalgebras that consist of elements of even ranks for even $n$.
(These cases aren't in the 1)-12) items of the theorem because we get reducible subalgebras.)

\end{theorem}

Now we want (just as it has been made in the Theorem 6) to find some subgroups of pseudo-unitary Lie group $\Wcl^\C(p,q)$ such that Lie algebras from Theorem 13 correspond to these Lie groups. There are 12 types of these Lie algebras (see previous theorem). It can be easily checked that for Clifford algebra of the sufficiently big dimension $n$ there are 31 subalgebras for even $n$ and 43 subalgebras for odd $n$ (we mean subalgebras in the form of the direct sums of subspaces of the given ranks). Let write down some subgroups of Lie group $\Wcl^\C(p,q)$ and Lie algebras that correspond to these Lie groups for Clifford algebras of small dimensions $n$.

$$n=1$$
$$\begin{tabular}{|c|c|}\hline
Lie algebra & Lie group\\\hline
$i\widehat{\textbf{0}}\qquad$&$\{\exp(i\varphi e), \varphi\in\R\}=\{(\cos\varphi)e+(i\sin\varphi)e, \varphi\in\R\}$\\\hline
$i\widehat{\textbf{1}}\qquad$&$\{\exp(i\varphi e^1), \varphi\in\R\}=\left\lbrace
\begin{array}{ll}
\{(\cos\varphi)e+(i\sin\varphi)e^1, \varphi\in\R\}, & \mbox{\rm $(p,q)=(1,0)$;}\\
\{(\ch\varphi)e+(i\sh\varphi)e^1, \varphi\in\R\}, & \mbox{\rm $(p,q)=(0,1)$}
\end{array}
\right.$\\\hline
\end{tabular}$$

$$n=2$$
$$\begin{tabular}{|c|c|}\hline
Lie algebra & Lie group\\\hline
$i\widehat{\textbf{0}}\qquad$&$\{\exp(i\varphi e), \varphi\in\R\}=\{(\cos\varphi)e+(i\sin\varphi)e, \varphi\in\R\}$\\\hline
$\widehat{\textbf{2}}\qquad$&$\{\exp(\varphi e^{12}), \varphi\in\R\}=\left\lbrace
\begin{array}{ll}
\{(\cos\varphi)e+(i\sin\varphi)e^{12}, \varphi\in\R\}, & \mbox{\rm $(p,q)=(2,0)$;}\\
\{(\ch\varphi)e+(i\sh\varphi)e^{12}, \varphi\in\R\}, & \mbox{\rm $(p,q)=(1,1)$;}\\
\{(\cos\varphi)e+(i\sin\varphi)e^{12}, \varphi\in\R\}, & \mbox{\rm $(p,q)=(0,2)$}
\end{array}
\right.$\\\hline
$i\widehat{\textbf{1}}\oplus\widehat{\textbf{2}}\qquad$&$\{U\in\widehat{\textbf{0}}\oplus\widehat{\textbf{2}}\oplus i\widehat{\textbf{1}}: U^* U=e\}$\\\hline
$i\widehat{\textbf{0}}\oplus\widehat{\textbf{2}}\qquad$&$\{U\in\widehat{\textbf{0}}\oplus\widehat{\textbf{2}}\oplus i\widehat{\textbf{0}}\oplus i\widehat{\textbf{2}}=\cl^\C_{even}(p,q): U^* U=e\}$\\\hline
\end{tabular}$$

$$n=3$$
$$\begin{tabular}{|c|c|}\hline
Lie algebra & Lie group\\\hline
$i\widehat{\textbf{0}}\qquad$&$\{\exp(i\varphi e), \varphi\in\R\}=\{(\cos\varphi)e+(i\sin\varphi)e, \varphi\in\R\}$\\\hline
$\widehat{\textbf{3}}\qquad$&$\{\exp(\varphi e^{123}), \varphi\in\R\}$\\\hline
$i\widehat{\textbf{1}}\oplus\widehat{\textbf{2}}\qquad$&$\{U\in\widehat{\textbf{0}}\oplus\widehat{\textbf{2}}\oplus i\widehat{\textbf{1}}\oplus i\widehat{\textbf{3}}: U^* U=e\}$\\\hline
$i\widehat{\textbf{0}}\oplus\widehat{\textbf{2}}\qquad$&$\{U\in\widehat{\textbf{0}}\oplus\widehat{\textbf{2}}\oplus i\widehat{\textbf{0}}\oplus i\widehat{\textbf{2}}=\cl^\C_{even}(p,q): U^* U=e\}$\\\hline
$\widehat{\textbf{2}}\qquad$&$\{U\in\widehat{\textbf{0}}\oplus\widehat{\textbf{2}}=\cl^\R_{even}(p,q): U^* U=e\}$\\\hline
$\widehat{\textbf{2}}\oplus\widehat{\textbf{3}}\qquad$&$\{U\in\widehat{\textbf{0}}\oplus\widehat{\textbf{1}}\oplus \widehat{\textbf{2}}\oplus \widehat{\textbf{3}}=\cl^\R(p,q): U^* U=e\}$\\\hline
$i\widehat{\textbf{1}}\oplus\widehat{\textbf{2}}\oplus\widehat{\textbf{3}}\qquad$&$\{U\in\cl^\C(p,q):\det\footnotemark U=1,\, U^* U=e\}$\\\hline
\end{tabular}$$
\footnotetext{Hereinafter determinant of Clifford algebra element is determinant of any of its matrix representation of minimal dimension. See \cite{Marchuk:Shirokov}.}
$$n=4$$
$$\begin{tabular}{|c|c|}\hline
Lie algebra & Lie group\\\hline
$i\widehat{\textbf{0}}\qquad$&$\{\exp(i\varphi e), \varphi\in\R\}=\{(\cos\varphi)e+(i\sin\varphi)e, \varphi\in\R\}$\\\hline
$i\widehat{\textbf{4}}\qquad$&$\{\exp(i \varphi e^{1234}), \varphi\in\R\}$\\\hline
$i\widehat{\textbf{1}}\oplus\widehat{\textbf{2}}\qquad$&$\{U\in\cl^\R_{even}(p,q)\oplus i\cl^\R_{odd}(p,q): U^* U=e\}$\\\hline
$\widehat{\textbf{2}}\qquad$&$\{U\in\widehat{\textbf{0}}\oplus\widehat{\textbf{2}}\oplus\widehat{\textbf{4}}=\cl^\R_{even}(p,q): U^* U=e\}$\\\hline
$\widehat{\textbf{2}}\oplus\widehat{\textbf{3}}\qquad$&$\{U\in\widehat{\textbf{0}}\oplus\widehat{\textbf{1}}\oplus \widehat{\textbf{2}}\oplus \widehat{\textbf{3}}\oplus\widehat{\textbf{4}}=\cl^\R(p,q): U^* U=e\}$\\\hline
$i\widehat{\textbf{1}}\oplus\widehat{\textbf{2}}\oplus\widehat{\textbf{3}}\oplus i\widehat{\textbf{4}}\qquad$&$\{U\in\cl^\C(p,q):\det U=1,\, U^* U=e\}$\\\hline
$i\widehat{\textbf{0}}\oplus\widehat{\textbf{2}}\oplus i\widehat{\textbf{4}}\qquad$&$\{U\in\cl^\C_{even}(p,q): U^* U=e\}$\\\hline
$\widehat{\textbf{2}}\oplus i\widehat{\textbf{4}}\qquad$&$\{U\in\cl^\C_{even}(p,q):\det U=1,\, U^* U=e\}$\\\hline
\end{tabular}$$

\bigskip
The proof of these statements is similar to the proof of Theorem 6.

Note that Lie algebra $i\widehat{\textbf{0}}$ correspond to Lie group $\{\exp(i\varphi e), \varphi\in\R\}=\{(\cos\varphi)e+(i\sin\varphi)e, \varphi\in\R\}$ for Clifford algebra $\cl^\C(p,q)$ of any dimension $n$ because Lie algebra $u(1)=\{i\varphi, \varphi\in\R\}$ correspond to unitary Lie group $U(1)=\{\exp(i\varphi), \varphi\in\R\}$.

\bigskip

\noindent{\bf Acknowledgements.} The author is grateful to
N.~G.~Marchuk for the constant attention to this work.

\end{document}